\RequirePackage{tikz}
\documentclass[doublecol]{epl2} 

\usepackage{amsfonts}
\usepackage{amssymb}
\usepackage{mathtools}
\usepackage[english]{babel}
\usepackage{graphicx}
\usepackage{subcaption}
\usepackage{pstricks}
\usepackage{psfrag}

\addto\captionsenglish{\renewcommand{\figurename}{Fig.}}

\title{Mapping the jamming transition of bidisperse mixtures}
\shorttitle{} 

\author{D.J.~Koeze \and D.~V{\aa}gberg \and B.B.T.~Tjoa \and B.P.~Tighe}
\shortauthor{D.J.~Koeze \etal}

\institute{                    
Process \& Energy Institute \\
Leeghwaterstraat 39, 2628 CB Delft \\
Delft University of Technology
}


\newcommand{\phic}{\phi_\mathrm{c}}
\newcommand{\fr}{f_\mathrm{r}}
\newcommand{\fn}{f_\mathrm{n}}
\newcommand{\fj}{f_\mathrm{j}}
\newcommand{\rhos}{\rho_\mathrm{s}}
\newcommand{\rhol}{\rho_\mathrm{l}}
\newcommand{\Tt}{\Delta_\mathrm{t}}
\newcommand{\Ts}{\Delta_\mathrm{s}}
\newcommand{\Tl}{\Delta_\mathrm{l}}
\newcommand{\Tm}{\Delta_\mathrm{m}}
\newcommand{\ie}{\textit{i.e.}}
\newcommand{\eg}{\textit{e.g.}}
\newcommand{\viz}{\textit{viz.}}

\graphicspath{{figs/}}

\abstract{
We systematically map out the jamming transition of all 2D bidisperse mixtures of frictionless disks in the hard particle limit. The critical volume fraction, mean coordination number, number of rattlers,  structural order parameters, and bulk modulus  each show a rich variation with mixture composition and particle size ratio, and can therefore  be tuned by choosing certain mixtures. We identify two local minima in the critical volume fraction, both of which have low structural order; one minimum is close to the widely studied 50:50 mixture of particles with a ratio of radii of 1:1.4. We also identify a region at low size ratios characterized by increased structural order and high rattler fractions, with a corresponding enhancement in the stiffness.
}

\begin{document}

\maketitle

\section{Introduction}
Soft materials such as foams, emulsions, and granulates exhibit a nonequilibrium fluid-solid, or ``jamming'', transition when their density is increased.  While most jammed matter is three-dimensional, jammed structures also occur in 2D when particles are trapped at interfaces, \eg~in bubble rafts, Pickering emulsions and ice flows. Numerous scaling exponents characterizing geometry and mechanics are also found to be the same in 2D and 3D \cite{goodrich2014,vanhecke2010}. As simulations in 2D are also (comparatively) computationally inexpensive, studies of jammed disk packings feature prominently in the literature \cite{goodrich2014,vanhecke2010,ohern2003,vagberg2011,vagberg2014,boschan2016,vandeen2014,vagberg2011a,luding2001,ueda2011}.

\begin{figure}
 \center
 \begin{subfigure}{.155\textwidth}
  \includegraphics*[width=\textwidth]{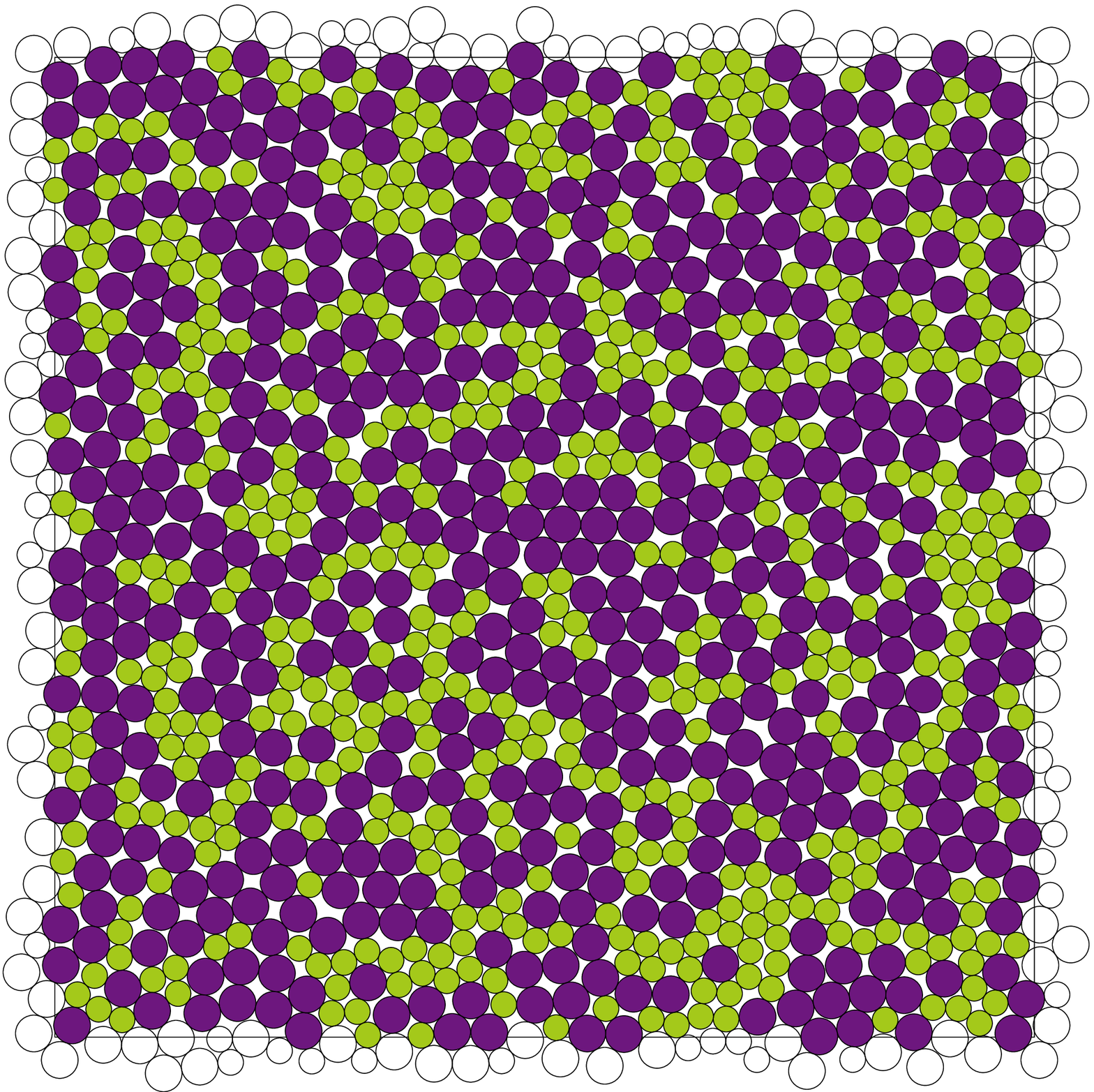}
  \caption{}
  \label{fig:packings:a}
 \end{subfigure}
 \begin{subfigure}{.155\textwidth}
  \includegraphics*[width=\textwidth]{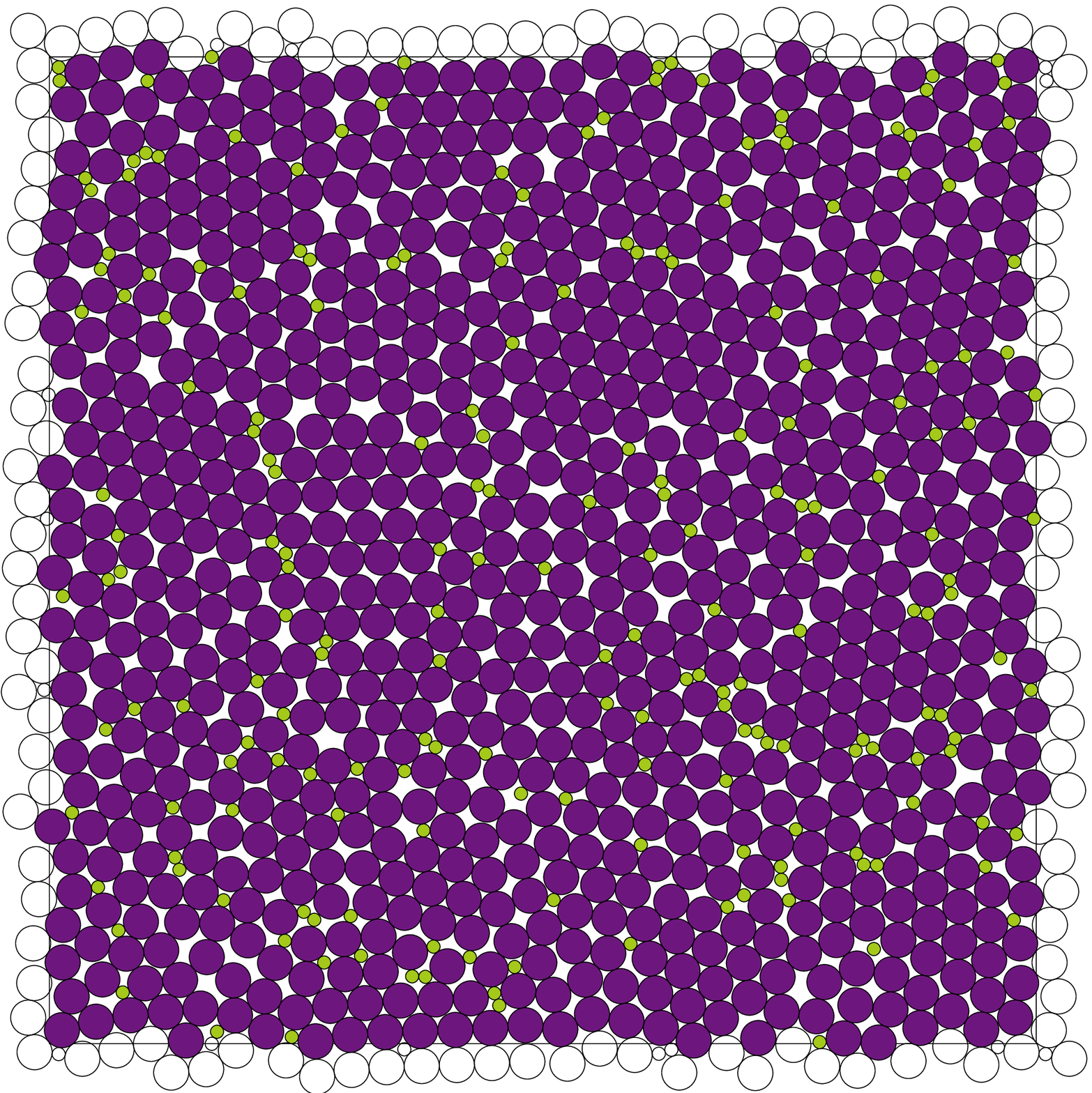}
  \caption{}
  \label{fig:packings:b}
 \end{subfigure}
 \begin{subfigure}{.155\textwidth}
  \includegraphics*[width=\textwidth]{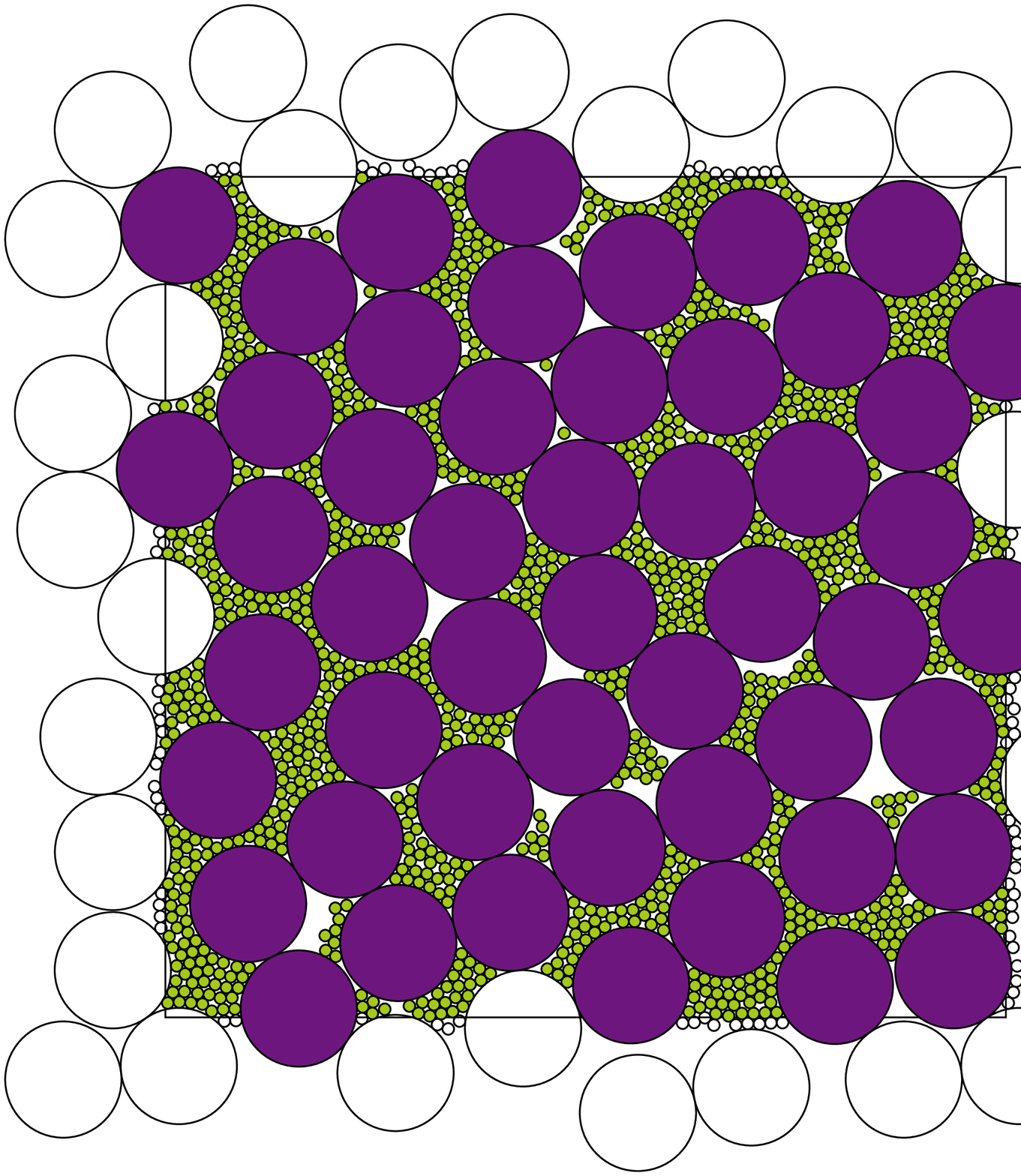}
  \caption{}
  \label{fig:packings:c}
 \end{subfigure}
 \begin{subfigure}{.155\textwidth}
  \includegraphics*[width=\textwidth]{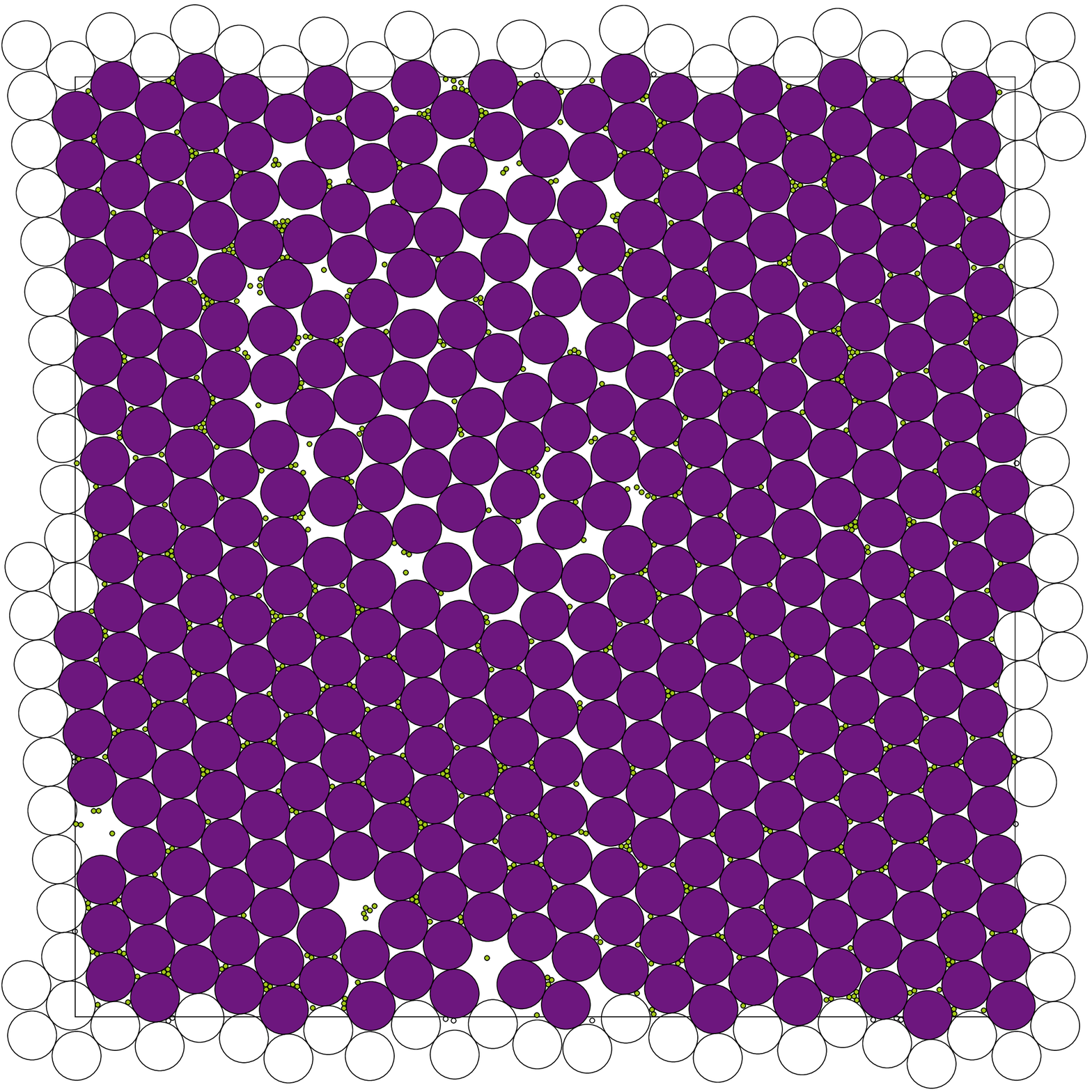}
  \caption{}
  \label{fig:packings:d}
 \end{subfigure}
 \begin{subfigure}{.155\textwidth}
  \includegraphics*[width=\textwidth]{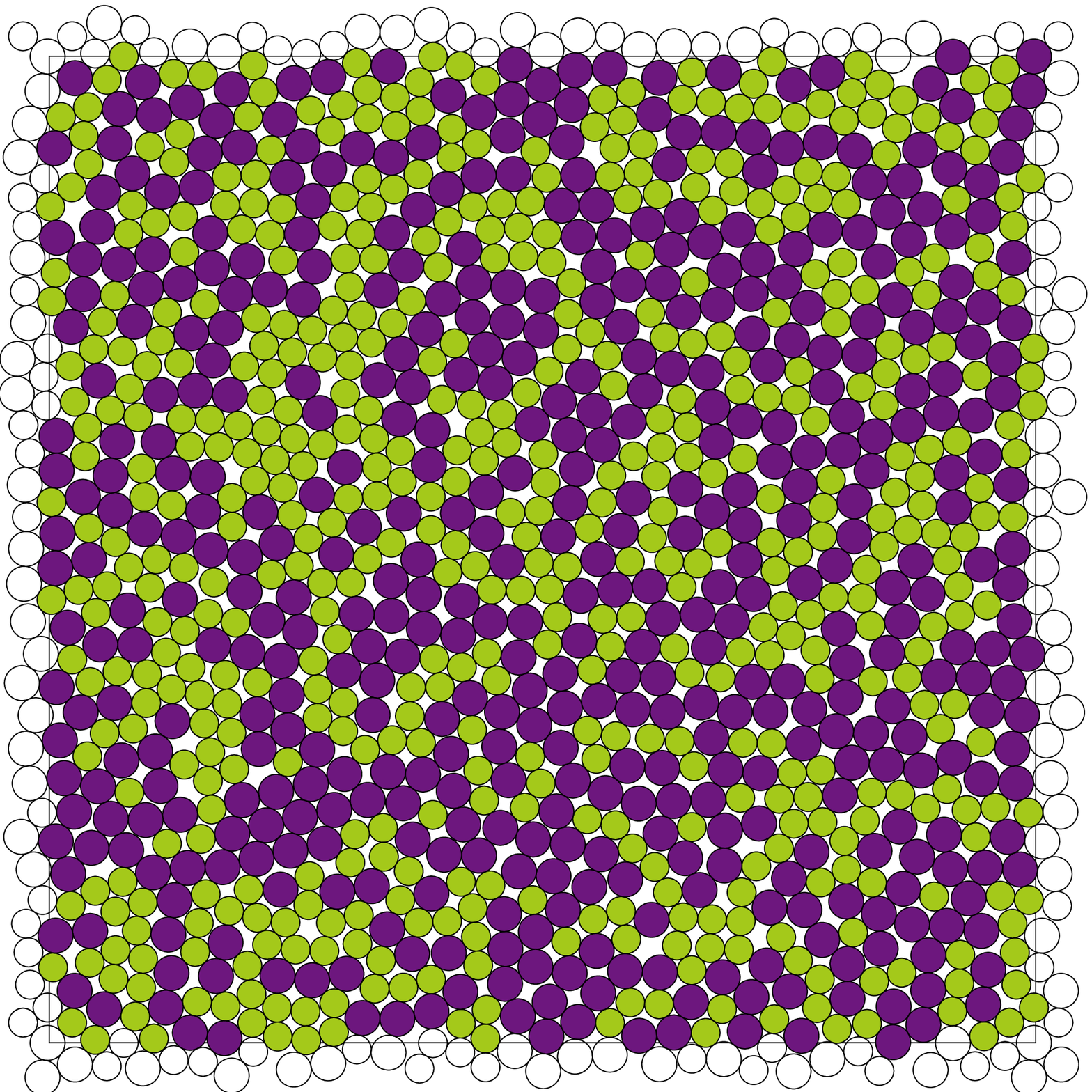}
  \caption{}
  \label{fig:packings:e}
 \end{subfigure}
 \begin{subfigure}{.155\textwidth}
  \includegraphics*[width=\textwidth]{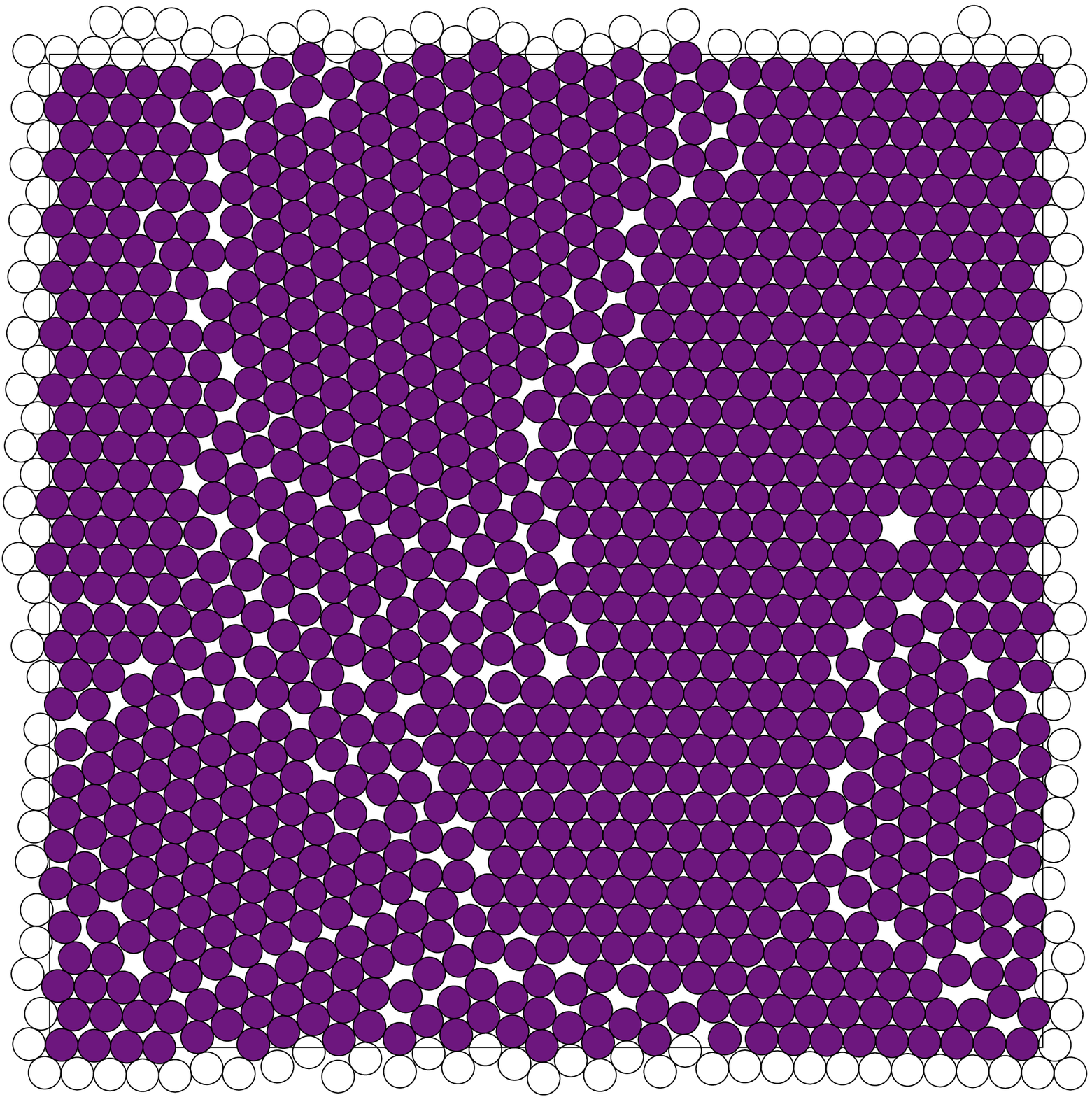}
  \caption{}
  \label{fig:packings:f}
 \end{subfigure}
 \caption{Bidisperse disk packings taken from the correspondingly labeled points in fig.~\ref{fig:phic:contour}.}
 \label{fig:packings}
\end{figure}

Disorder is a distinguishing feature of jammed packings \cite{bernal1959,bierwagen1974,nelson1982,nelson1984,rubinstein1982}, and indeed the critical volume fraction $\phic$ where frictionless packings jam is often denoted random close packing (RCP), although this name is controversial \cite{torquato2000}. Unlike spheres, rapidly quenched monodisperse disk packings tend to crystallize -- hence bidisperse mixtures are widely favored (Fig.~\ref{fig:packings}). By convention, most studies consider the `classic' mixture of small and large disks -- a 50:50 ratio of the number of disks with a 1:1.4 size ratio of their radii \cite{ohern2003,perera1999,speedy1999} -- selected for their low degree of structural order, see \eg~\cite{goodrich2014,vanhecke2010,ohern2003,perera1999,speedy1999,vagberg2011,vagberg2014,boschan2016,vandeen2014,vagberg2011a}. Far less is known about other number and size ratios \cite{donev2004,kumar2015,luding2001,ogarko2012,ueda2011,zhao2012,zhao2014,puckett2011}. The goal of the present work is to systematically map the space of bidisperse frictionless hard disk mixtures for the first time. 

We have several main findings. By varying the number ratio $\fn$ and size ratio $\fr$, we find rich structure in $\phic(\fr,\fn)$, including a large region with low $\phic$ and low structural order appropriate to studies of jamming. This region includes the `classic' mixture, which is close to the global minimum of $\phic$. However there is a second local minimum which, to our knowledge, has not previously been reported; this minimum may be of particular interest for future jamming studies. We find that $\phic$ increases dramatically for mixtures with extreme difference in their radii, as small particles fill in the voids formed by large particles. These and other mixtures exhibit local crystalline order and a high number of non-load bearing particles (rattlers), which both correlate strongly and positively with the bulk modulus. Hence it may be possible to use mixture composition and size ratio to tune stiffness.

\section{Model and methods}
Bidisperse mixtures contain $N_\mathrm{a}$ particles with radius $R_\mathrm{a}$ and $N_\mathrm{b}$ particles with radius $R_\mathrm{b}$. Packings are characterized by the number fraction ${\fn=N_\mathrm{a} / (N_\mathrm{a} + N_\mathrm{b})}$ and size ratio ${\fr = R_\mathrm{a}/R_\mathrm{b}}$. The space of all bidisperse mixtures is then given by ${0 \leq \fn \leq 1}$ and ${0 \leq \fr \leq \infty}$. On five lines in this domain the mixture is monodisperse: ${\fr=0}$, ${\fr=1}$, ${\fr=\infty}$, ${\fn=0}$ and ${\fn=1}$. Furthermore, any point $(\fr,\fn)$ in the domain represents  the same mixture as the point ${(1/\fr, 1-\fn)}$, except for a change of scale. We therefore consider only mixtures where ${\fr \leq 1}$. The remaining $(\fr,\fn)$ parameter space is sampled with a regular $41 \times 41$ grid.

Our simulations approach $\phic$ from the jammed phase, using soft particles with a finite ranged repulsive potential 
\begin{eqnarray}
U_{ij} = 
\begin{cases}
({k}/{2}) \left( R_i + R_j - r_{ij} \right)^2 & \quad R_i+R_j \geq r_{ij} \\
0 & \quad R_i + R_j < r_{ij} \, .
\end{cases}
\label{eq:potential}
\end{eqnarray}
Here $k$ is the spring constant and sets the energy scale, $r_{ij}$ is the center-to-center distance between particles $i$ and $j$, and $R_i$ is the radius of particle $i$. The unit of length is set by the radius of the larger particle, $R_\mathrm{b} = 1$.

\begin{figure}
 \center
 \psfrag{phi}{$\phi$}
 \psfrag{fj}{$\fj$}
 \psfrag{1}{\footnotesize $1$}
 \psfrag{0.8}{\footnotesize $0.8$}
 \psfrag{0.6}{\footnotesize $0.6$}
 \psfrag{0.4}{\footnotesize $0.4$}
 \psfrag{0.2}{\footnotesize $0.2$}
 \psfrag{0}{\footnotesize $0$}
 \psfrag{0.835}{\footnotesize $0.835$}
 \psfrag{0.84}{\footnotesize $0.840$}
 \psfrag{0.845}{\footnotesize $0.845$}
 \psfrag{0.85}{\footnotesize $0.850$}
 \psfrag{0.855}{\footnotesize $0.855$}
 \psfrag{a2}{\scriptsize $0.2$}
 \psfrag{a3}{\scriptsize $0.4$}
 \psfrag{a4}{\scriptsize $0.6$}
 \psfrag{a5}{\scriptsize $1$}
 \psfrag{0.839}{\scriptsize $0.839$}
 \psfrag{0.840}{\scriptsize $0.840$}
 \includegraphics*[width=.4\textwidth]{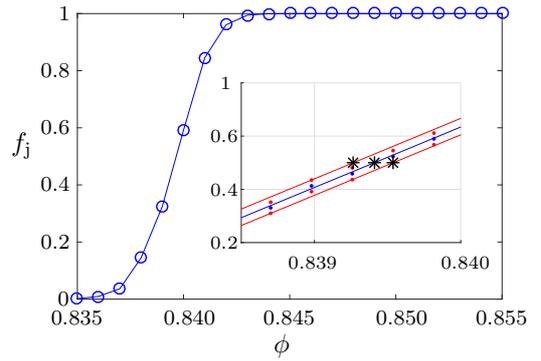}
 \caption{Fraction of jammed packings $\fj$ at different packing fractions $\phi$, for $\fr=0.7$ and $\fn=0.5$. (inset) $\phic$ is estimated with linear regression on five points in the linear part of this curve. The stars indicate the estimate for $\phic$ and the error bounds.}
 \label{fig:fj-vs-phi}
\end{figure}

We use 2D packings of $N=N_\mathrm{a} + N_\mathrm{b} = 1024$ particles (unless stated otherwise) generated by instantaneously quenching an infinite temperature configuration to a local energy minimum using the nonlinear conjugate gradient method of \cite{vagberg2011}. Packings are created at a specified $\phi$, yielding an ensemble with a distribution of pressures \cite{vagberg2014}. The criterion for jamming is also the same as in \cite{vagberg2011}, \viz~the energy is larger than a low threshold. Fig.~\ref{fig:fj-vs-phi} shows the fraction of jammed packings $\fj$ versus the packing fraction $\phi$. The location of the jamming transition for a certain $\fr$ and $\fn$ is taken as the packing fraction $\phic$ where half of the packings are jammed, \ie~$\fj(\phic)=0.5$. We estimate $\phic$ from a regression through five points close to ${\fj=0.5}$, see fig.~\ref{fig:fj-vs-phi} (inset).  At every $\phi$ there is a statistical uncertainty associated with the binomial distribution of finding a jammed packing. The resulting uncertainty is at most $0.006$ in $\phic$, estimated with the largest deviation from the regression line.

\begin{figure}[h!]
 \center
 \begin{subfigure}{.49\textwidth}
  \psfrag{fr}{$\fr$}
  \psfrag{fn}{$\fn$}
  \psfrag{phic}{$\phic$}
  \psfrag{1}{\small $1$}
  \psfrag{0.8}{\small $0.8$}
  \psfrag{0.6}{\small $0.6$}
  \psfrag{0.4}{\small $0.4$}
  \psfrag{0.2}{\small $0.2$}
  \psfrag{0}{\small $0$}
  \psfrag{0.82}{\small $0.82$}
  \psfrag{0.84}{\small $0.84$}
  \psfrag{0.86}{\small $0.86$}
  \psfrag{0.88}{\small $0.88$}
  \psfrag{0.90}{\small $0.90$}
  \psfrag{0.92}{\small $0.92$}
  \psfrag{0.94}{\small $0.94$}
  \includegraphics*[width=\textwidth]{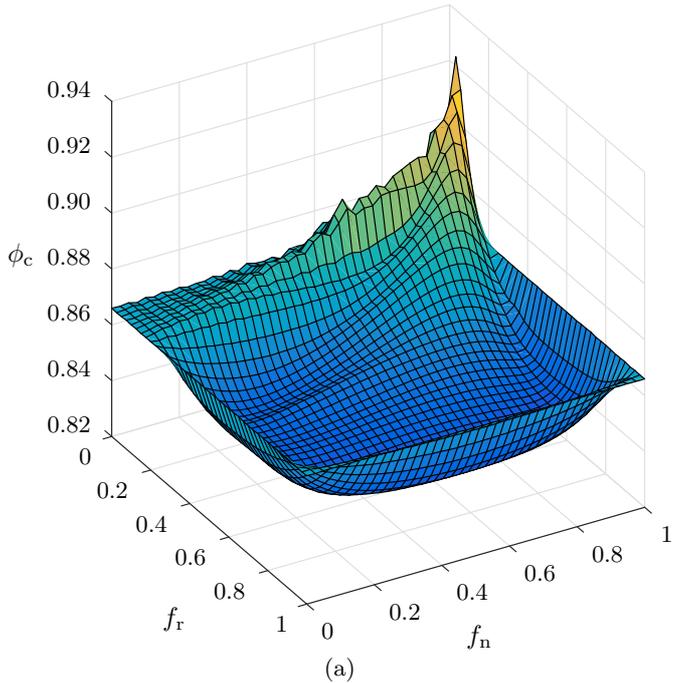}
  \caption{}
  \label{fig:phic:surface}
 \end{subfigure}
 \begin{subfigure}{.49\textwidth}
  \psfrag{fr}{$\fr$}
  \psfrag{fn}{$\fn$}
  \psfrag{1}{\small $1$}
  \psfrag{0.8}{\small $0.8$}
  \psfrag{0.6}{\small $0.6$}
  \psfrag{0.4}{\small $0.4$}
  \psfrag{0.2}{\small $0.2$}
  \psfrag{0}{\small $0$}
  \psfrag{a1}{$\mathbf{a}$}
  \psfrag{b1}{$\mathbf{b}$}
  \psfrag{c1}{$\mathbf{c}$}
  \psfrag{d1}{$\mathbf{d}$}
  \psfrag{e1}{$\mathbf{e}$}
  \psfrag{f1}{$\mathbf{f}$}
  \psfrag{a2}{$\mathbf{a'}$}
  \psfrag{b2}{$\mathbf{b'}$}
  \psfrag{c2}{$\mathbf{c'}$}
  \psfrag{d2}{$\mathbf{d'}$}
  \psfrag{e2}{$\mathbf{e'}$}
  \psfrag{f2}{$\mathbf{f'}$}
  \psfrag{fs}{$\fr^*$}
  \psfrag{v1}{$2$}
  \psfrag{v2}{$1$}
  \psfrag{v3}{$\frac{1}{2}$}
  \psfrag{v4}{$\frac{1}{4}$}
  \psfrag{v5}{$\frac{1}{8}$}
  \psfrag{v6}{\hspace{-.7mm}$\frac{1}{16}$}
  \includegraphics*[width=\textwidth]{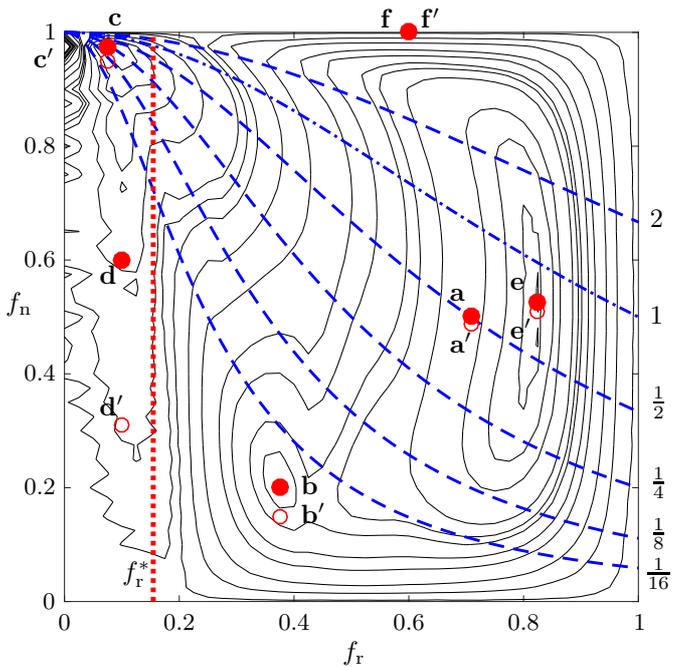}
  \caption{}
  \label{fig:phic:contour}
 \end{subfigure}
 \caption{(a) $\phic$ surface as a function of number and size ratio. (b) Contours of the same surface. The (dot-)dashed curves follow constant ratio of species volume (value indicated on the right). Voids in a triangle of large particles can hold at least one small rattler for $\fr < \fr^*$ (the dotted line). Mixtures that are discussed are marked (see fig.~\ref{fig:packings}); where a prime denotes the shift of the point when using the corrected $\fn'$.}
 \label{fig:phic}
\end{figure}

\section{Jamming of mixtures}

We find rich structure in the location of the jamming transition $\phic(\fr,\fn)$, as depicted in both surface and contour plots  in fig.~\ref{fig:phic}. Several features stand out: flat lines along the domain edges;  a clear peak in $\phic$ in the upper corner, with a corresponding ridge; and two basins around local minima, labeled $\mathbf{b}$ and $\mathbf{e}$. We discuss each feature in turn.

{\em Flat edges.--} The edges of the domain correspond to monodisperse mixtures. Hence $\phic$ is constant, though not that of a triangular lattice due to defects, the cause of which we discuss in more detail later. Along the edge ${\fr=0}$ we find stronger finite size effects, since $R_a = 0$ and so the effective number of particles changes with $\fn$. Near this edge minimizing the energy is numerically challenging due to strong separation of scales in the two radii. In this region there is a slight systematic overestimate of the jammed fraction $\fj$ for packings close to jamming, resulting in a small underestimate of $\phic$. 

The critical volume fraction falls off sharply upon moving away from three of the four domain edges, which suggests even weak bidispersity inhibits structural order.  We find radically different behavior moving away from the edge $\fr=0$. Here $\phic$ first increases, hinting that packings at low $\fr$ are qualitatively different in their structure.

\begin{figure}
 \center
 \begin{subfigure}{.24\textwidth}
  \psfrag{fr}{\small $\fr$}
  \psfrag{fn}{\small $\fn$}
  \psfrag{rho}{\small $\rhos$}
  \psfrag{1}{\scriptsize $1$}
  \psfrag{0.8}{\scriptsize $0.8$}
  \psfrag{0.6}{\scriptsize $0.6$}
  \psfrag{0.4}{\scriptsize $0.4$}
  \psfrag{0.2}{\scriptsize $0.2$}
  \psfrag{0}{\scriptsize $0$}
  \psfrag{0.5}{\scriptsize $0.5$}
  \includegraphics*[width=\textwidth]{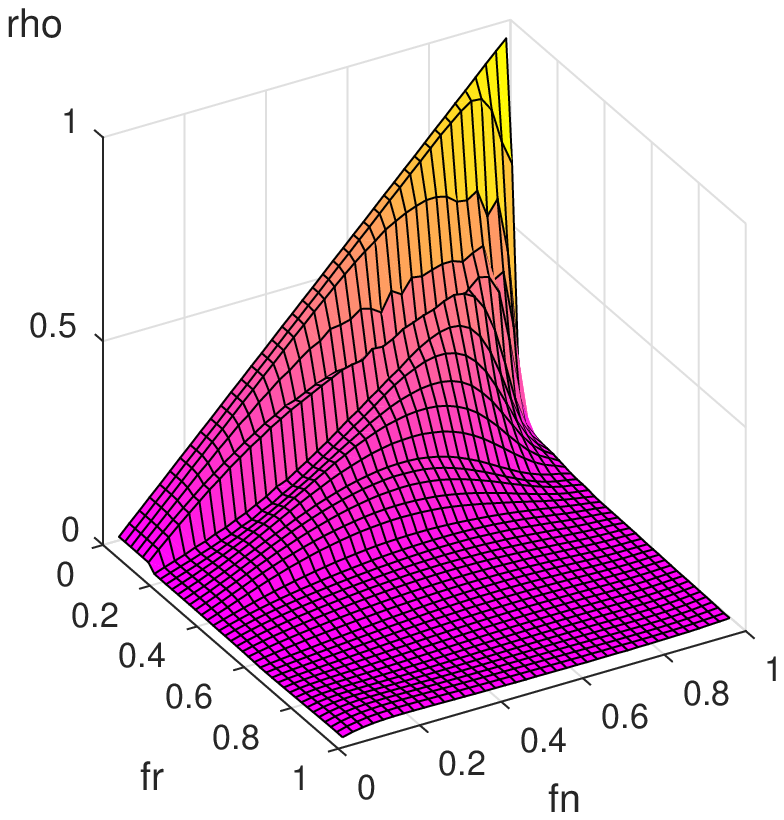}
  \caption{Small species}
  \label{fig:rattlers:small}
 \end{subfigure}
 \begin{subfigure}{.24\textwidth}
  \psfrag{fr}{\small $\fr$}
  \psfrag{fn}{\small $\fn$}
  \psfrag{rho}{\small $\rhol$}
  \psfrag{1}{\scriptsize $1$}
  \psfrag{0.8}{\scriptsize $0.8$}
  \psfrag{0.6}{\scriptsize $0.6$}
  \psfrag{0.4}{\scriptsize $0.4$}
  \psfrag{0.2}{\scriptsize $0.2$}
  \psfrag{0}{\scriptsize $0$}
  \psfrag{0.02}{\scriptsize $0.02$}
  \psfrag{0.04}{\scriptsize $0.04$}
  \includegraphics*[width=\textwidth]{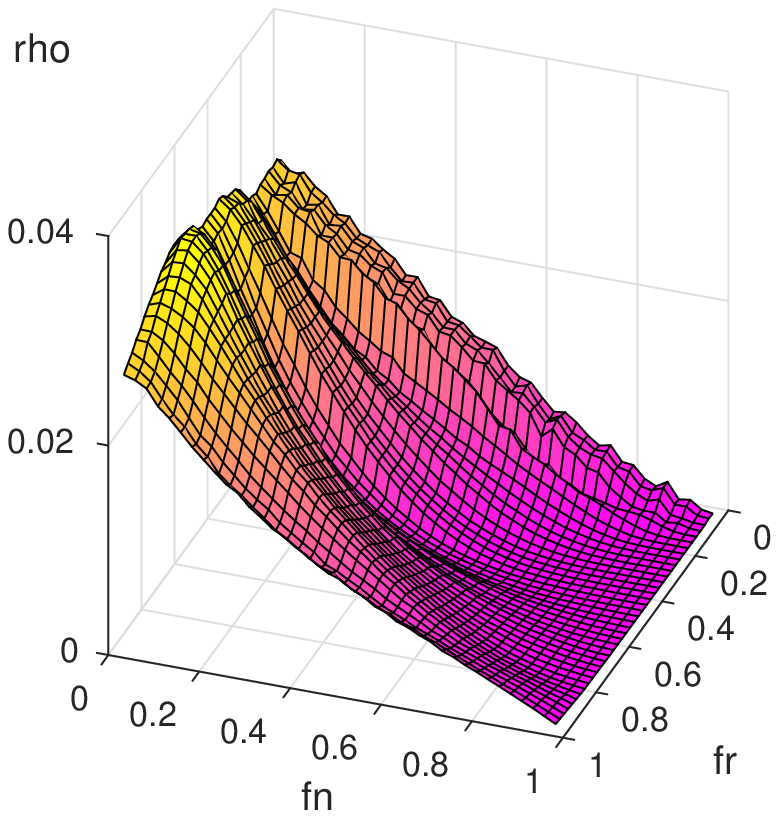}
  \caption{Large species}
  \label{fig:rattlers:large}
 \end{subfigure}
 \begin{subfigure}{.4\textwidth}
   \psfrag{Z}{\hspace{-3mm}\small $Z$}
   \psfrag{fr}{\small $\fr$}
   \psfrag{fn}{\small $\fn$}
   \psfrag{4}{\scriptsize\hspace{-2.5mm}$4.0$}
   \psfrag{4.5}{\hspace{-1mm}\scriptsize $4.5$}
   \psfrag{5}{\scriptsize\hspace{-2.4mm}$5.0$}
   \psfrag{0}{\scriptsize$0$}
   \psfrag{0.2}{\scriptsize$0.2$}
   \psfrag{0.4}{\scriptsize$0.4$}
   \psfrag{0.6}{\scriptsize$0.6$}
   \psfrag{0.8}{\scriptsize$0.8$}
   \psfrag{1}{\scriptsize$1$}
   \includegraphics*[width=\textwidth]{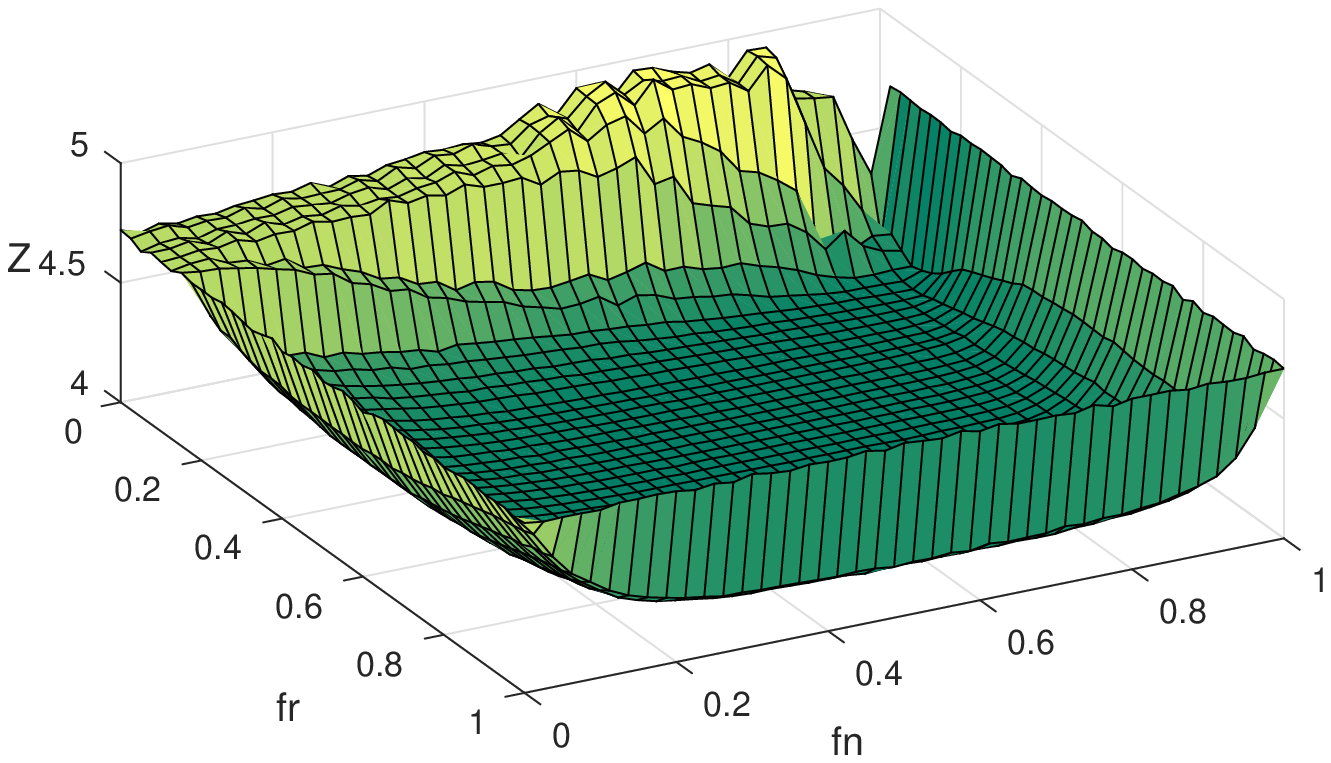}
   \caption{Coordination number}
   \label{fig:rattlers:coordination}
 \end{subfigure}
 \caption{(a) and (b) Fraction of small ($\rhos$) and large ($\rhol$) rattlers. Monodisperse edges are cut away; note the different orientation of (b). (c) Average coordination number $Z$ of jammed packings.}
 \label{fig:rattlers}
\end{figure}

{\em Ridge and peak.--} 
When ${\fr < \fr^* \equiv 2/\sqrt{3}-1}$ (dotted line in fig.~\ref{fig:phic:contour}) the small species fits within the void formed by a triangle of large particles (visible in figs.~\ref{fig:packings:c},\ref{fig:packings:d}). In this region many of the small particles are `rattlers' (they bear no load), hence are effectively not part of the contact network, as can be seen from the linear increase of small rattler fraction $\rhos$ in fig.~\ref{fig:rattlers:small}. Due to void filling, high critical packing fractions are reached along this edge, exceeding even ${\phi_\mathrm{xtal} \approx 0.9069}$ of the monodisperse triangular lattice. When following a line of fixed $\fr$ from small to large $\fn$ (corresponding to the ridge in $\phic$), we depart from a monodisperse packing of large particles by introducing rattlers in the (mostly triangular) voids formed by large particles. Filling continues until for some $\fn$ there are enough small particles that they frustrate the jammed structure of large particles and simultaneously become load-bearing particles -- see the sharp decrease in $\rhos$ in fig.~\ref{fig:rattlers:small}. 

When $\fr$ is small nearly all small particles are rattlers, and the load-bearing network is monodisperse. A defect-free triangular lattice of large particles can only form when the lattice is commensurate with the simulation box; incommensurate boxes generate defects, evident as fine structure in $\phic$ along the $\fn$-direction (see fig.~\ref{fig:phic:surface}).

{\em Two minima.--}
The commonly chosen mixture of ${\fr=1.4}$ and ${\fn=0.5}$ (marked \textbf{a} in fig.~\ref{fig:phic:contour}) sits in the larger of two `basins' in the $\phic$ surface. Close to point $\mathbf{a}$ we find the mixture with the lowest $\phic$, around ${\fr=0.8}$, ${\fn=0.5}$, labeled \textbf{e} in fig.~\ref{fig:phic:contour}. There is a second basin with its own local minimum around $\fr=0.4$, ${\fn=0.2}$, marked \textbf{b} in fig.~\ref{fig:phic:contour}. To our knowledge this second minimum has not previously been observed. In the following Section we present several measures indicating that packings in both basins have low structural order. One consequence is that these states are indeed appropriate for studies of jamming.

{\em Rattlers and coordination number.--} So far we have taken $\fn$ to be the fraction of small particles. While this is the relevant quantity when preparing packings experimentally, it is not for understanding the underlying load-bearing network. We have re-calculated the $\phic$ surface using a corrected $\fn$ -- denoted $\fn'$ -- which is the number fraction of only the non-rattler particles (figs.~\ref{fig:rattlers:small},\ref{fig:rattlers:large}). Using $\fn'$, the points $\mathbf{a}$ - $\mathbf{e}$ shift to the points labeled $\mathbf{a'}$ - $\mathbf{e'}$ ($\mathbf{f}$ and $\mathbf{f'}$ are the same, as it is monodisperse). The only significant shift is from $\mathbf{d}$ to $\mathbf{d'}$, which sits in the region $\fr < \fr^*$ where rattlers are most prevalent.

The mean coordination number $Z$ is a geometric measure with important consequences for mechanical response. \cite{vanhecke2010} Simple counting arguments dating to Maxwell predict the isostatic value $Z_{\rm iso} = 4$ in a disordered jammed disk mixture. Fig.~\ref{fig:rattlers:coordination} depicts the  coordination number averaged over the jammed states of our ensemble. Away from the monodisperse edges, there is a flat region in the $Z$ surface with a height ${Z \approx 4.1}$, consistent with Maxwell. The difference from 4 is because our ensemble has a spread in the distance to jamming of individual packings. In the region ${\fr < \fr^*}$ and for lower $\fn$ the $Z$ surface is at the same height as the monodisperse edge, as all small particles are rattlers. Going to higher $\fn$ we start seeing the effects of small particles jamming in the voids of the large particle crystal.

\section{Local order}
\begin{figure}
 \center
 \begin{subfigure}[b]{.24\textwidth}
  \psfrag{DD}{$\Tt$}
  \psfrag{fr}{\small $\fr$}
  \psfrag{fn}{\small $\fn$}
  \psfrag{1}{\scriptsize $1$}
  \psfrag{0.8}{\scriptsize $0.8$}
  \psfrag{0.6}{\scriptsize $0.6$}
  \psfrag{0.4}{\scriptsize $0.4$}
  \psfrag{0.2}{\scriptsize $0.2$}
  \psfrag{0}{\scriptsize $0$}
  \psfrag{2}{\scriptsize $2$}
  \psfrag{4}{\scriptsize $4$}
  \psfrag{6}{\scriptsize $6$}
  \psfrag{8}{\scriptsize $8$}
  \includegraphics*[width=\textwidth]{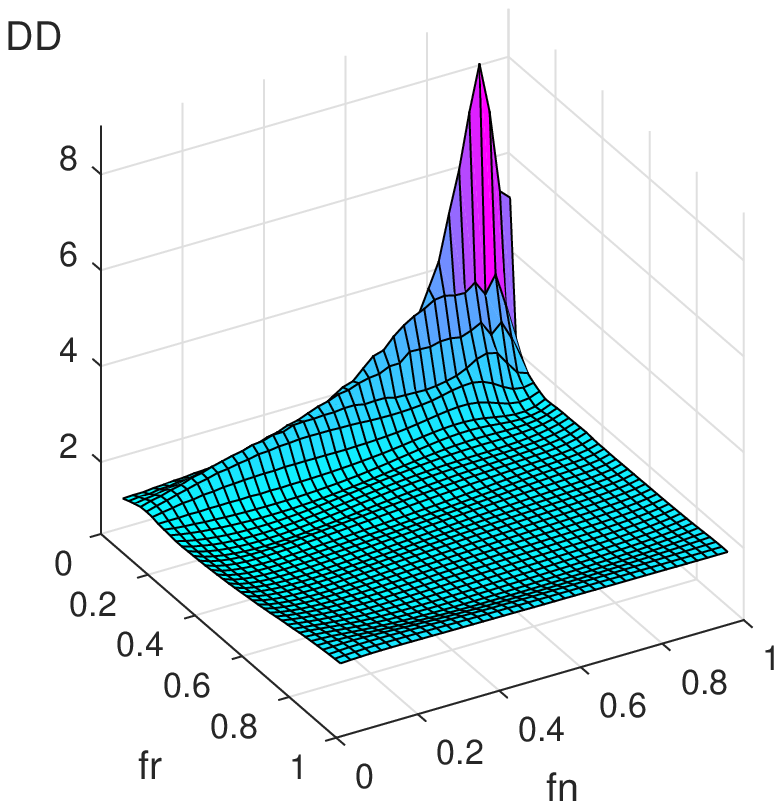}
  \caption{Total count}
  \label{fig:triangles:all}
 \end{subfigure}
 \begin{subfigure}[b]{.24\textwidth}
  \psfrag{DD}{$\Tl$}
  \psfrag{fr}{\small $\fr$}
  \psfrag{fn}{\small $\fn$}
  \psfrag{1}{\scriptsize $1$}
  \psfrag{0.8}{\scriptsize $0.8$}
  \psfrag{0.6}{\scriptsize $0.6$}
  \psfrag{0.4}{\scriptsize $0.4$}
  \psfrag{0.2}{\scriptsize $0.2$}
  \psfrag{0}{\scriptsize $0$}
  \psfrag{0.5}{\scriptsize $0.5$}
  \psfrag{1.5}{\scriptsize $1.5$}
  \psfrag{2}{\scriptsize $2$}
  \includegraphics*[width=\textwidth]{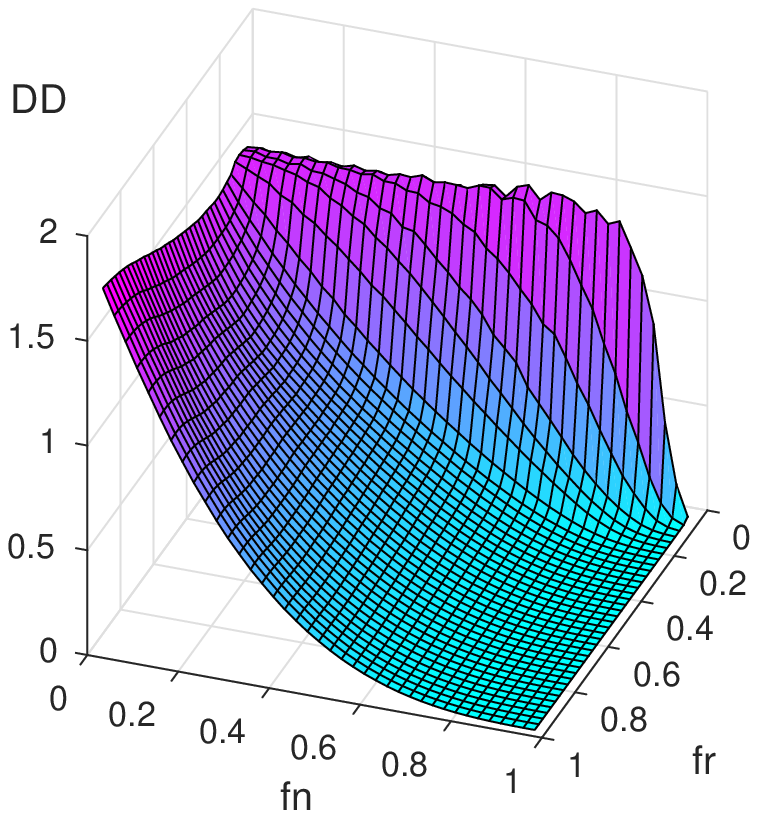}
  \caption{Large species}
  \label{fig:triangles:big}
 \end{subfigure}
 \begin{subfigure}{.24\textwidth}
  \psfrag{DD}{$\Ts$}
  \psfrag{fr}{\small $\fr$}
  \psfrag{fn}{\small $\fn$}
  \psfrag{1}{\scriptsize $1$}
  \psfrag{0.8}{\scriptsize $0.8$}
  \psfrag{0.6}{\scriptsize $0.6$}
  \psfrag{0.4}{\scriptsize $0.4$}
  \psfrag{0.2}{\scriptsize $0.2$}
  \psfrag{0}{\scriptsize $0$}
  \psfrag{1}{\scriptsize $1$}
  \psfrag{2}{\scriptsize $2$}
  \psfrag{3}{\scriptsize $3$}
  \includegraphics*[width=\textwidth]{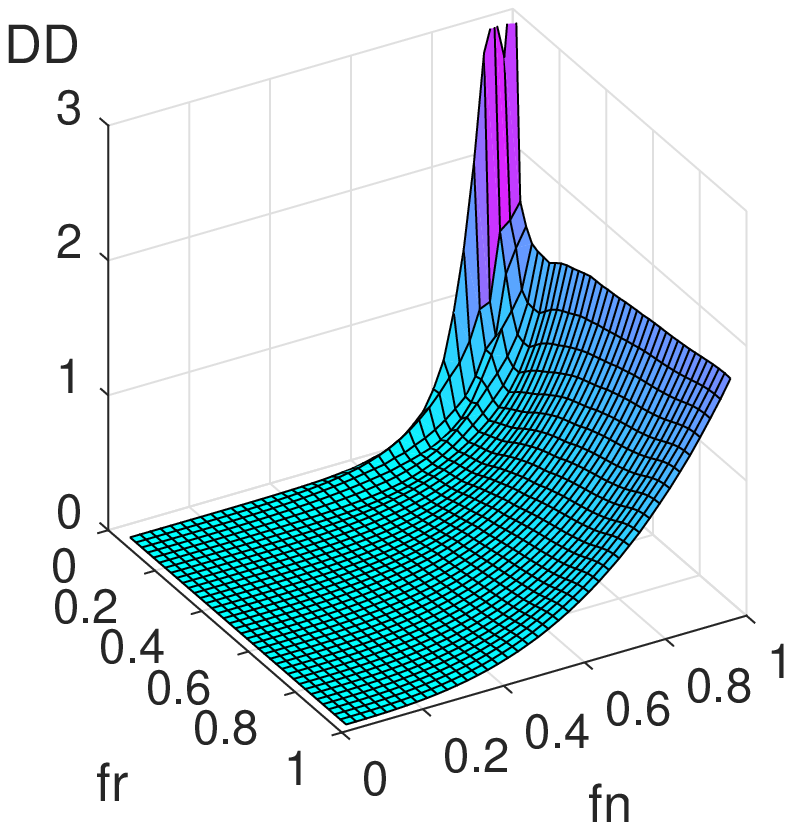}
  \caption{Small species}
  \label{fig:triangles:small}
 \end{subfigure}
 \begin{subfigure}{.24\textwidth}
  \psfrag{DD}{$\Tm$}
  \psfrag{fr}{\small $\fr$}
  \psfrag{fn}{\small $\fn$}
  \psfrag{1}{\scriptsize $1$}
  \psfrag{0.8}{\scriptsize $0.8$}
  \psfrag{0.6}{\scriptsize $0.6$}
  \psfrag{0.4}{\scriptsize $0.4$}
  \psfrag{0.2}{\scriptsize $0.2$}
  \psfrag{0}{\scriptsize $0$}
  \psfrag{2}{\scriptsize $2$}
  \psfrag{4}{\scriptsize $4$}
  \includegraphics*[width=\textwidth]{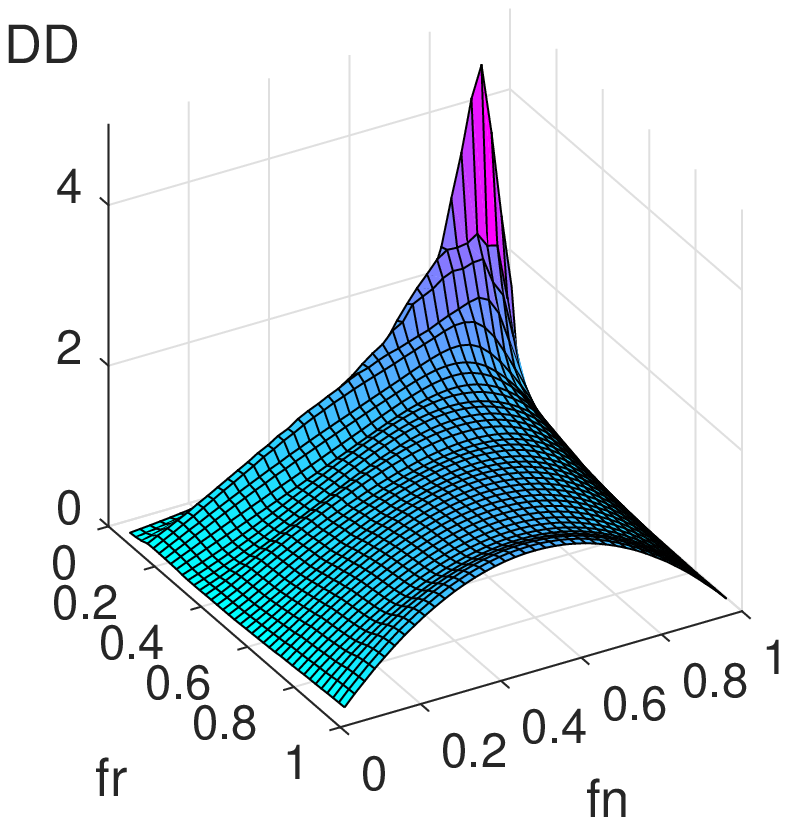}
  \caption{Mixed species}
  \label{fig:triangles:mix}
 \end{subfigure}
 \caption{The number of triangles per non-rattler particle, one measure of crystallinity. Note that the monodisperse edge is cut away and the orientation of (b) is different.}
 \label{fig:triangles}
\end{figure}

Order in a packing can be characterized by many different order parameters \cite{torquato2000}. In monodisperse packings we find large patches with triangular lattice structure, while in the basins relatively few triangles are present ({\em c.f.}~figs.~\ref{fig:packings:f} and \ref{fig:packings:a}). To characterize this local order, we count the number of triangles formed by three particles in contact normalized by the number of non-rattler particles in the packing. Figure \ref{fig:triangles:all} shows this measure for all mixtures. Observe that indeed the peak and ridge have a high count of triangles and are therefore relatively ordered. The basins, including the commonly used point $\fr = 1.4$ and $\fn=0.5$, indeed have a low count of triangles, indicating lower structural order.

Surprisingly, we find the fewest triangles per particle near the local minimum $\mathbf{b}$ rather than the global minimum $\mathbf{e}$, where $\phic$ is lower. Point $\mathbf{b}$ seems to be optimal for frustrating the triangular crystalline structure of the large particles -- which, recall, is the reason point $\mathbf{a}$ is commonly selected for simulations. Therefore point $\mathbf{b}$ could be an interesting alternative to the widely studied point $\mathbf{a}$. Packings at $\mathbf{b}$  differ from those at $\mathbf{a}$/$\mathbf{e}$ in that the small species particles are smaller and there are fewer of them. 

\hyphenation{par-ti-ci-pate} 

Even when triangles are present in the contact network, they can form a disordered tiling. Such disorder is introduced when many triangles are formed by particles of both species. In these packings four different shapes and sizes of triangles make up a large part of the contact network, resulting in highly disordered structures. Figure \ref{fig:triangles:mix} shows the number of `mixed' triangles in which both species participate, again normalized with the number of non-rattlers. There is a curve of maximal mixed triangles that passes through the global minimum of $\phic$. This suggests there is a qualitative difference in the disorder found in the local and global minima. 

\begin{figure}
 \center
 \psfrag{phi}{$\phic$}
 \psfrag{d}{$d$}
 \psfrag{0}{\small $0$}
 \psfrag{0.2}{\small $0.2$}
 \psfrag{0.4}{\small $0.4$}
 \psfrag{0.6}{\small $0.6$}
 \psfrag{0.8}{\small $0.8$}
 \psfrag{1}{\small $1$}
 \psfrag{0.82}{\small $0.82$}
 \psfrag{0.84}{\small $0.84$}
 \psfrag{0.86}{\small $0.86$}
 \psfrag{0.88}{\small $0.88$}
 \psfrag{0.90}{\small $0.90$}
 \psfrag{0.92}{\small $0.92$}
 \includegraphics*[width=.4\textwidth]{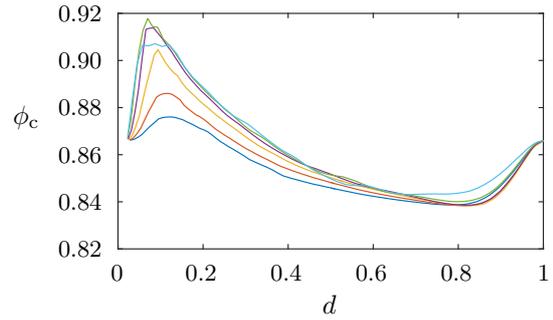}
 \caption{Along curves of constant volume ratio (dashed lines in fig.~\ref{fig:phic:contour}) $\phic$ has a similar form. $d$ is the relative distance along the curves.}
 \label{fig:ratio_lines}
\end{figure}

One way to rationalize the curve of maximal mixed triangles is inspired by the intuitive idea that it is most likely to find a particle of the other species as a neighbor when the volume occupied by the species is roughly equal. The ratio between the volumes can be written as $V_\mathrm{a}/V_\mathrm{b} = \fr^2 \fn/(1-\fn)$. For any such ratio we find curves of the form seen in fig.~\ref{fig:phic:contour}; the dot-dashed curve, in particular, indicates mixtures for which the ratio is unity. This curve closely follows the ridge in the number of mixed triangles. We further note that the curves of constant volume ratio converge around the peak. Plotting $\phic$ along the curves produces similar profiles with the peak and minimum in roughly the same location, see fig.~\ref{fig:ratio_lines}. 

\section{Stiffness}
States below the jamming transition are not rigid and cannot bear load, hence their stiffness or bulk modulus $K$ is zero. However, when the jamming transition is approached from above $K$ tends to a positive value \cite{ohern2003}. Here we present $K$ for all bidisperse mixtures,  computed by applying a small volumetric strain to an ensemble of initially jammed packings and minimizing the energy. From the change in pressure $p$ we determine the  modulus as $K = (1/\phi) (\mathrm{d}p / \mathrm{d}\phi)$. As in experiments, a small number of contact changes can occur during the compression step, but numerical tests indicate that the ensemble-averaged modulus is unaffected \cite{boschan2016,vandeen2014}. 

\begin{figure}
 \center
 \psfrag{fr}{$\fr$}
 \psfrag{fn}{$\fn$}
 \psfrag{K}{$K$}
 \psfrag{1}{\small $1$}
 \psfrag{0.8}{\small $0.8$}
 \psfrag{0.7}{\small $0.7$}
 \psfrag{0.6}{\small $0.6$}
 \psfrag{0.5}{\small $0.5$}
 \psfrag{0.4}{\small $0.4$}
 \psfrag{0.3}{\small $0.3$}
 \psfrag{0.2}{\small $0.2$}
 \psfrag{0}{\small $0$}
 \includegraphics*[width=.45\textwidth]{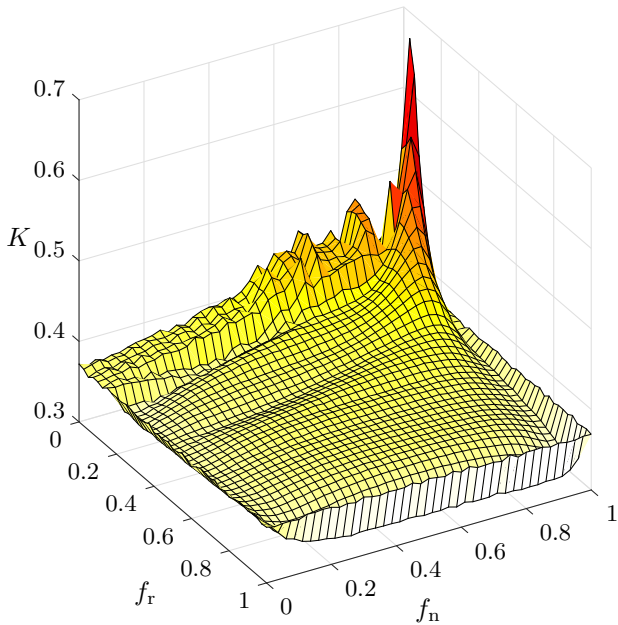}
 \caption{Bulk modulus for all $\fr$ and $\fn$. 
 }
 \label{fig:bulk}
\end{figure}

We plot the bulk modulus surface in fig.~\ref{fig:bulk}. Packings on the monodisperse edge have the same stiffness, as expected. States in the interior where $\fr > \fr^*$ show little difference from the monodisperse mixture. However there is a sharp increase in $K$ in the region where small particles fit inside voids of triangles of touching large particles. The stiffest packings occur for low $\fr$ and high $\fn$.

\begin{figure}
\center
 \psfrag{fr}{$\fr$}
 \psfrag{phi}{\hspace{-4mm} $\phic$}
 \psfrag{K}{\hspace{-4.3mm} $K$}
 \psfrag{0.9}{\small $0.9$}
 \psfrag{0.92}{\small $0.92$}
 \psfrag{0.94}{\small $0.94$}
 \psfrag{0.96}{\small $0.96$}
 \psfrag{0.98}{\small $0.98$}
 \psfrag{1}{\small $1$}
 \psfrag{0.4}{\hspace{-3.8mm} \small $0.40$}
 \psfrag{0.35}{\hspace{-2.5mm} \small $0.35$}
 \psfrag{0.3}{\hspace{-3.8mm} \small $0.30$}
 \psfrag{0.25}{\hspace{-2.5mm} \small $0.25$}
 \psfrag{0.84}{\hspace{-2.5mm} \small $0.84$}
 \psfrag{0.85}{\hspace{-2.5mm} \small $0.85$}
 \psfrag{0.86}{\hspace{-2.5mm} \small $0.86$}
 \psfrag{0.87}{\hspace{-2.5mm} \small $0.87$}
 \psfrag{512}{\scriptsize $512$}
 \psfrag{1024}{\scriptsize $1024$}
 \psfrag{2048}{\scriptsize $2048$}
 \psfrag{4096}{\scriptsize $4096$}
 \includegraphics*[width=.49\textwidth]{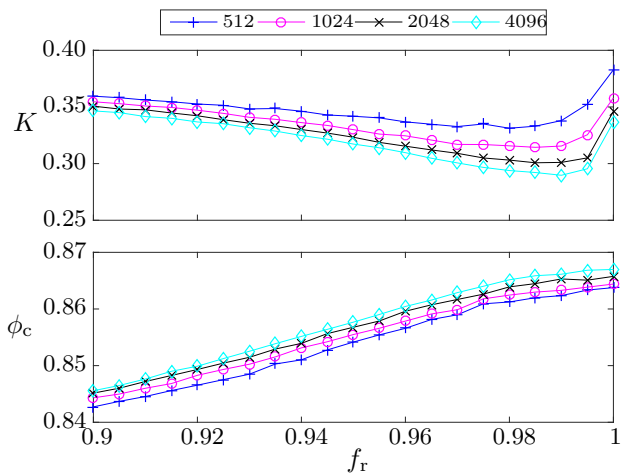}
\caption{A cross section of the critical packing fraction $\phic$ and the bulk modulus $B$ near the monodisperse edge. This cross section is for a fixed value of $\fn = 0.5$.}
\label{fig:edge_limit}
\end{figure}

At the monodisperse limit there is a sharp increase in the stiffness. To determine if the increase is discontinuous, we take smaller steps towards this limit in a cross section with fixed ${\fr=0.6}$. While true discontinuities can only be found in the large system size limit, we find that the increase in $K$ becomes sharper with increasing N (see fig.~\ref{fig:edge_limit}). This indeed suggests there is a discontinuity in the bulk modulus $K$ at the edge, while the critical packing fraction $\phic$ goes to the monodisperse limit continuously.

We caution that one must be careful when comparing the moduli and stresses of systems with different bidisperse mixtures, as varying $\fr$ and $\fn$ has two distinct effects. The first, which we probe in fig.~\ref{fig:bulk}, is the collective effect of structural differences in the contact network due to changes in the mixture composition and particle size ratio. The second effect is due to the fact that stress carries units of energy per unit volume, and is therefore sensitive both to an overall rescaling of the particle sizes, and to differences in box volume for different $(\fr,\fn)$ at fixed $N$. The second effect can be quite large, as there is a factor 40 difference in the volume of a packing in the peak and one in the opposite corner of the $\phic$ landscape. 

The natural units of stress are given by $k_{\rm eff}/R^{D-2}$, where $R$ is the typical particle size and the effective spring constant $k_{\rm eff} \equiv f/\delta$ is determined by the ratio of the typical force $f$ on a contact and the typical dimensionful overlap $\delta$ between particles. For disks interacting via the potential in eq.~\ref{eq:potential} the effective spring constant is simply $k$ and the quantity $R^{D-2}$ is simply unity, and hence the natural units of stress are constant. However for other values of the spatial dimension $D$, and/or other forms of the pair potential, the quantity $k_{\rm eff}/R^{D-2}$ will generally be sensitive to changes in $R$ and/or the box volume. This is the case, \eg~, for the widely-used potential \cite{vanhecke2010,ohern2003,vagberg2011,vagberg2014,vagberg2011a,perera1999}
\begin{eqnarray}
{U_{ij} = \frac{\epsilon}{2} \left(1-\frac{r_{ij}}{R_i+R_j}\right)^2} \,,
\label{eq:rel_potential}
\end{eqnarray} 
where $\epsilon$ is a constant with units of energy and the quantity in parentheses is the dimensionless overlap. 

\begin{figure}[t!]
 \center
 \psfrag{phic}{$\phic$}
 \psfrag{fn}{$\fn$}
 \psfrag{128}{\scriptsize $128$}
 \psfrag{256}{\scriptsize $256$}
 \psfrag{512}{\scriptsize $512$}
 \psfrag{1024}{\scriptsize $1024$}
 \psfrag{0}{\small $0$}
 \psfrag{0.2}{\small $0.2$}
 \psfrag{0.4}{\small $0.4$}
 \psfrag{0.6}{\small $0.6$}
 \psfrag{0.8}{\small $0.8$}
 \psfrag{1}{\small $1$}
 \psfrag{0.83}{\small $0.83$}
 \psfrag{0.84}{\small $0.84$}
 \psfrag{0.85}{\small $0.85$}
 \psfrag{0.86}{\small $0.86$}
 \includegraphics*[width=.49\textwidth]{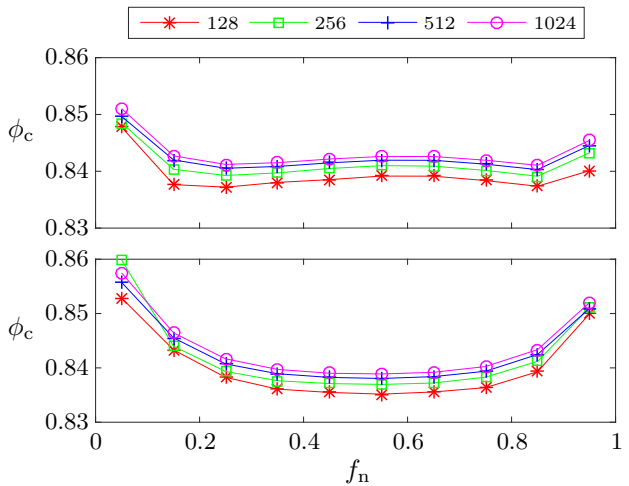}
 \caption{Finite size effects for two cross sections with fixed $\fr$ at $0.45$ and $0.65$.}
 \label{fig:finsize:crosssection}
\end{figure}

\begin{figure}
 \center
 
 \psfrag{phi}{\raisebox{3mm}{$\phic/\phic^{(8192)}$}}
 \psfrag{N}{$N$}
 \psfrag{128 }{\hspace{.7mm} \small $128$}
 \psfrag{256 }{\small $256$}
 \psfrag{512 }{\small $512$}
 \psfrag{1024}{\small $1024$}
 \psfrag{2048}{\small $2048$}
 \psfrag{4096}{\small $4096$}
 \psfrag{8192}{\small $8192$}
 \psfrag{0.96}{\hspace{-2.5mm} \small $0.96$}
 \psfrag{0.97}{\hspace{-2.5mm} \small $0.97$}
 \psfrag{0.98}{\hspace{-2.5mm} \small $0.98$}
 \psfrag{0.99}{\hspace{-2.5mm} \small $0.99$}
 \psfrag{1.00}{\hspace{-2.5mm} \small $1.00$}
 \psfrag{1.01}{\hspace{-2.5mm} \small $1.01$}
 \psfrag{a}{\scriptsize a}
 \psfrag{b}{\scriptsize b}
 \psfrag{c}{\scriptsize c}
 \psfrag{d}{\scriptsize d}
 \psfrag{e}{\scriptsize e}
 \psfrag{f}{\scriptsize f}
 \includegraphics*[width=.49\textwidth]{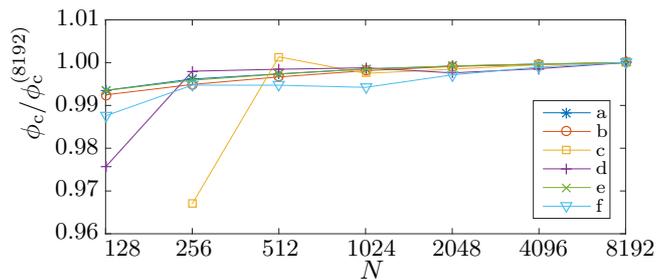}
 \caption{Finite size effects for the selected points in fig.~\ref{fig:phic:contour}.}
 \label{fig:finsize:points}
\end{figure}

\section{Finite size effects}
The structure in $\phic$ remains the same in the infinite system limit. Figure \ref{fig:finsize:crosssection} shows $\phic$ for two cross sections in the parameter space. We see that the cross sections uniformly increase in $\phic$ with increasing $N$. This shift converges with $N$, as apparent for the selected points from fig.~\ref{fig:phic:contour} plotted in fig.~\ref{fig:finsize:points}. This means that the surface shifts upwards slightly, while retaining the features, \ie~the peak, two minima, the ridge and the monodisperse edges. For an extensive study of these finite-size effects for a certain mixture see \eg~\cite{goodrich2014}.

We furthermore see that the fraction $\fj$ of jammed packings has a sigmoidal shape (see fig.~\ref{fig:fj-vs-phi}) that tends to a step function with increasing $N$. \cite{ohern2003,vagberg2011a} Since the surface only shifts we find that the sigmoidal shape of $\fj$ goes to a step function with increasing $N$ in the same way for all $(\fr,\fn)$.

\section{Conclusions}

We have mapped out the critical packing fraction of all mixtures of bidisperse packings. We find that there is large region in the $(\fr, \fn)$ parameter space that avoids structural order, and is therefore suitable for studies of jamming. In this sense the classic mixture of $\fr=1.4(\equiv 0.71)$ and $\fn=0.5$ is not special, which is a fortunate finding given that most numerical studies address just this one point in state space. Intriguingly, we find not one but two local minima in the $\phic$ surface. While the global minimum is close to the classic mixture, the other minimum is at lower $\fr$ and $\fn$ (${\fr=0.4}$ and ${\fn=0.2}$); we suggest that it might provide a useful alternate to the classic mixture for future numerical studies.

The region at low $\fr$, where small particles fit in voids formed by larger ones, is distinguished by a high rattler fraction, increased structural order, and, especially at high $\fn$, enhanced stiffness. Mixtures from this region should not be selected as `typical' disordered states. However, it may be possible to tune the packing stiffness by varying $\fn$ in this region. 

There is a number of ways to generalize our results. 3D sphere packings are an obvious, albeit computationally expensive, direction for future work; others include friction, particle shape, and other forms of polydispersity. It would also be worthwhile to verify the existence of the other minimum in $\phic$ experimentally (see also \cite{puckett2011}). Systems that are close to our model include a monolayer of foam and trapped emulsions. \cite{katgert2010,desmond2013}

\acknowledgments
We thank A. Brouwer, J. Brouwer and J. van Lochem for their contribution in the initial phase of this work. We also thank S. Luding for helpful discussions. DJK, DV and BPT acknowledge funding from the Netherlands Organization for Scientific Research (NWO). This work was sponsored by NWO Exacte Wetenschappen (Physical Sciences) for the use of supercomputer facilities.


\end{document}